\documentclass[aps,prl,twocolumn,amsmath,amsfonts,amssymb,floatfix,superscriptaddress]{revtex4-2}
\usepackage{epsfig,subfigure}
\usepackage{bm}
\usepackage{bbm}
\usepackage{color,soul,xcolor}
\usepackage{hyperref}
\usepackage{wrapfig}
\usepackage{lipsum}
\hypersetup{colorlinks=false,linkcolor=black}
\usepackage[left]{lineno}
\begin{document}

\title{Electronic and Magnonic Properties of $g$-Wave Altermagnetism in Intercalated Transition Metal Dichalcogenides}
\author{Shuyi Li}
\affiliation{Department of Physics, University of Florida, Gainesville, FL 32611, USA}

\author{Adrian Bahri}
\affiliation{Department of Physics, University of Florida, Gainesville, FL 32611, USA}

\author{Chunjing Jia}
\affiliation{Department of Physics, University of Florida, Gainesville, FL 32611, USA}

\date{\today}

\begin{abstract}
Altermagnetism is a recently identified class of magnetic order characterized by unconventional momentum-dependent spin splitting in the absence of net magnetization, and understanding its electronic and magnetic properties is essential for revealing its fundamental physics and potential applications. In this work we investigate two intercalated transition-metal dichalcogenides, Fe$_{1/4}$NbS$_2$ and V$_{1/3}$NbS$_2$, as candidate altermagnetic materials by using effective tight-binding and spin models complemented by first-principles calculations. We show that the $g$-wave electronic spin splitting originates from bond-dependent hopping anisotropy, leading to material-dependent nodal structures. For the magnetic excitations, the emergence of chiral splitting in the magnon dispersion is controlled by single-ion anisotropy, which manifests as altermagnetic-like nodal structures when spins are oriented along an easy-axis. Conversely, this altermagnetic signature disappears when the spins are aligned in an easy-plane. Beyond linear spin-wave theory, we find that $1/S$ corrections from magnon--magnon interactions preserve the symmetry and nodal structure of the band splitting while generally reducing its magnitude, with strong antiferromagnetic exchange leading to a non-negligible renormalization of the chiral splitting. Our findings establish intercalated transition-metal dichalcogenides as promising platforms for understanding the interplay between crystal symmetry, non-relativistic spin splitting, and magnetic properties in altermagnets.
\end{abstract}

\maketitle

\section{Introduction}
Recently, altermagnets have attracted significant attention due to their unconventional behavior, which distinguishes them from conventional collinear ferromagnets and antiferromagnets~\cite{AMth1,AMth2,AMth3,AMliu,AMth4,AMth5,AMth6,AMsplit1,AMsplit2}. In this class of magnetic order, magnetic moments form an antiferromagnetic-like arrangement with vanishing net magnetization, while the electronic structure exhibits a spin-dependent band splitting analogous to that of ferromagnets. This unique combination of compensated magnetic order and momentum-dependent spin splitting has been associated with a variety of interesting phenomena and potential applications, including distinct variants of the Hall effect~\cite{AMhall1,AMhall2,AMhall3,AMhall4,AMhall5,AMhall6,AMhall7,AMweyl1,AMweyl2}, proposed unconventional superconducting states~\cite{AMSC1,AMSC2,AMSC3,AMSC4,AMSC5,AMSC6,AMSC7,AMSC8}, and spintronic functionalities~\cite{AMth2,spintronics1,spintronics2}.

In addition to electronic structure, altermagnetic symmetry also manifests in magnetic excitations. Recent theoretical and experimental studies have shown that altermagnetism can give rise to unconventional magnon spectra, where modes with opposite chirality are split even in the absence of an external magnetic field~\cite{magnon1, magnon2, magnon3, magnon4, vb_am1}. Importantly, recent studies suggest that the nature of the magnon splitting in altermagnetic systems can depend sensitively on magnetic anisotropy. In particular, easy-axis systems tend to exhibit altermagnetic-like chiral splitting, whereas in easy-plane configurations the splitting may deviate from the characteristic altermagnetic form~\cite{vb_am1}. Furthermore, in antiferromagnetic exchange dominated altermagnets, quantum fluctuations arising from magnon--magnon interactions can significantly modify magnetic excitations, resulting in renormalization of the magnon spectrum and the emergence of spontaneous magnon decay~\cite{magnonint1, magnonint2, magnonint3}.

Motivated by these considerations, here we focus on magnetically intercalated transition-metal dichalcogenides $T_yMX_2$, which provide a promising platform for realizing altermagnetism. In these systems, magnetic ions intercalated in layered $MX_2$ hosts form ordered superstructures at typical concentrations $y = 1/3$ and $1/4$, giving rise to triangular lattices and antiferromagnetic order that naturally supports $g$-wave altermagnetism~\cite{Fe, V}. 

In this work, we focus on Fe$_{1/4}$NbS$_2$ and V$_{1/3}$NbS$_2$ as model platforms to investigate how lattice structure, anisotropy, and interactions influence altermagnetic electronic and magnonic properties. Guided by symmetry analysis and atomic arrangements, we developed effective tight-binding models that utilize bond-dependent hopping anisotropy to accurately capture $g$-wave altermagnetic spin splitting. The resulting material-dependent nodal structures are also confirmed by our first-principles calculations. Regarding magnetic excitations, we investigate how magnetic anisotropy influences the symmetry and momentum dependence of the chiral magnon splitting. Furthermore, we systematically investigate the effects of magnon--magnon interactions up to order $1/S$ within nonlinear spin-wave theory. For both material systems, these higher-order corrections renormalize the magnitude of the splitting while preserving its symmetry. Quantitatively, the $1/S$ corrections generally reduce the chiral splitting and become significant in antiferromagnetic exchange-dominated altermagnets. Our results establish the microscopic mechanism of $g$-wave altermagnetism in intercalated transition-metal dichalcogenides and elucidate how magnon chiral splitting depends on single-ion anisotropy and quantum fluctuations, providing guidance for future experimental studies.

\section{Crystal Structure and Magnetic Ground State}

\begin{figure}[htbp]
    \centering
    \includegraphics[width=0.9\linewidth]{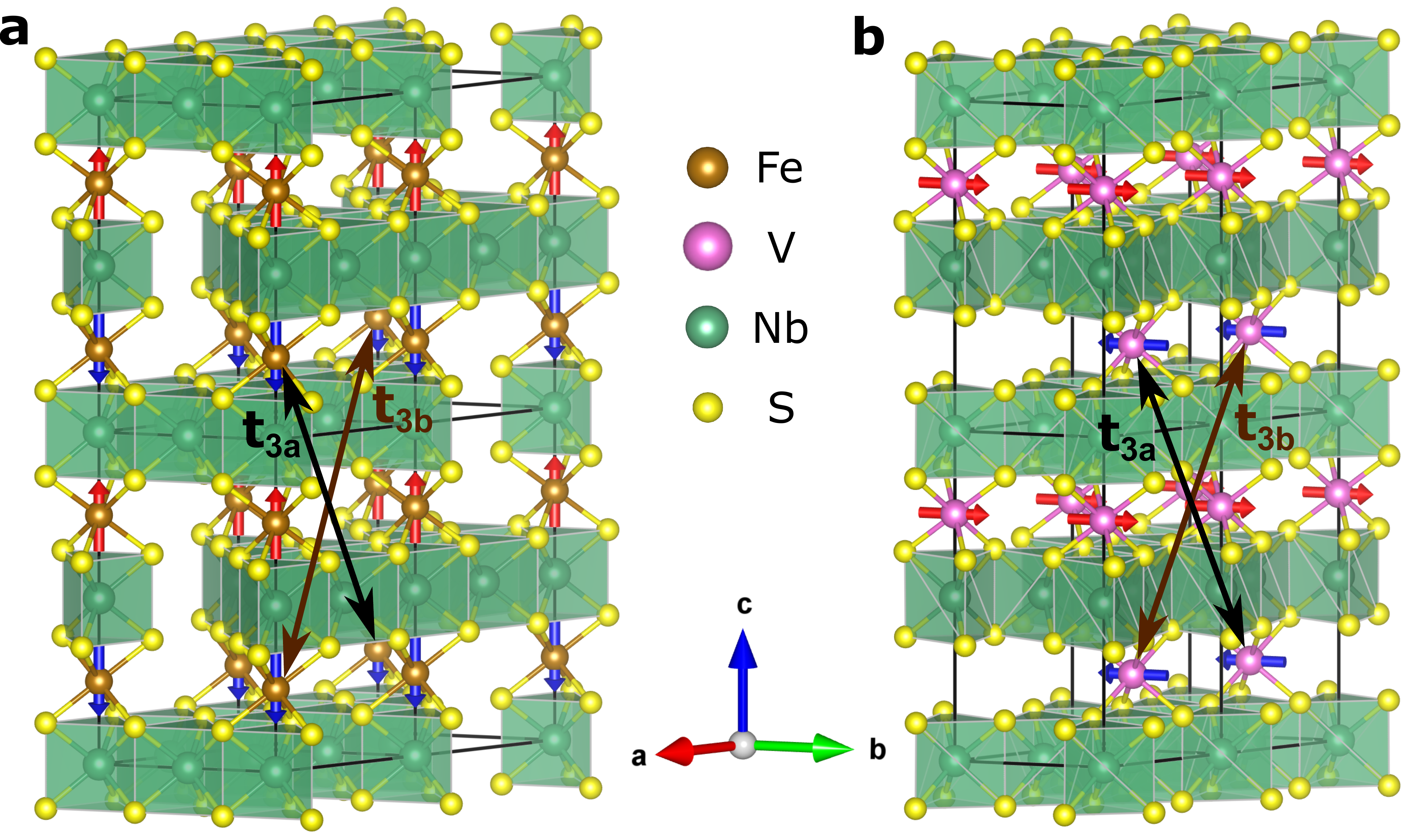}
    \caption{\textbf{Crystal structure and magnetic ground states of Fe$_{1/4}$NbS$_2$ and V$_{1/3}$NbS$_2$.} \textbf{a} Crystal structure and magnetic configuration of Fe$_{1/4}$NbS$_2$. \textbf{b} Crystal structure and magnetic configuration of V$_{1/3}$NbS$_2$. Brown, pink, green, and yellow spheres represent Fe, V, Nb, and S ions, respectively. Red and blue arrows penetrating the Fe and V ions indicate the local magnetic moments on those sites. The black and brown double arrows highlight the anisotropic hopping amplitudes $t_{3a}$ and $t_{3b}$, respectively.}
    \label{fig:lattice}
\end{figure}

The two materials considered in this work, Fe$_{1/4}$NbS$_2$ and V$_{1/3}$NbS$_2$, belong to the family of layered transition-metal dichalcogenides (TMDs) with space group $P6_3/mmc$ and $P6_3 22$. Their crystal structures are shown in Fig.~\ref{fig:lattice}. In both materials, NbS$_2$ forms quasi-two-dimensional layers stacked along the crystallographic $c$ axis. Each NbS$_2$ layer consists of edge-sharing NbS$_6$ octahedra arranged on a triangular lattice. Between adjacent NbS$_2$ layers, the intercalated transition-metal ions form triangular lattices, realizing a $2\times2$ centrosymmetric superlattice in Fe$_{1/4}$NbS$_2$ and a $\sqrt{3}\times\sqrt{3}$ noncentrosymmetric superlattice in V$_{1/3}$NbS$_2$.

Both materials exhibit antiferromagnetic arrangement of the localized moments on the transition-metal ions between the NbS$_2$ layers. The magnetic ground state of either material is characterized by ferromagnetic alignment within each layer and antiferromagnetic stacking along the $c$ direction. From the perspective of spin-group symmetry, the presence of the non-magnetic NbS$_2$ layers breaks the combined symmetry $\{C_{2\perp}||\mathcal{P}\bm{\tau}\}$ while preserving $\{C_{2\perp}||C_{6z}\bm{\tau}\}$, where $C_{2\perp}$ denotes a $180^\circ$ rotation around an axis perpendicular to the spin direction, $C_{6z}$ denotes a $60^\circ$ rotation around the $z$($c$) axis, $\mathcal{P}$ is inversion, and $\bm{\tau}$ represents translation. This symmetry configuration places both materials in the class of $g$-wave altermagnets. Furthermore, in Fe$_{1/4}$NbS$_2$ (V$_{1/3}$NbS$_2$), the localized moments are oriented out of plane (in plane), reflecting easy-axis (easy-plane) magnetic anisotropy. In the following discussion, the magnetic unit-cell vectors are chosen as $(a,0,0)$, $(-a/2,\sqrt{3}a/2,0)$, and $(0,0,c)$, where $a$ and $c$ are the corresponding lattice constants.

\section{Electronic Structure from Tight-Binding Model}

\subsection{Effective Hamiltonian}
To model the electronic structure of these two altermagnetic systems, 
we consider the effective tight-binding Hamiltonian in Eq.~(\ref{ELE_H}), 
which includes only the magnetic sites,
\begin{equation}\label{ELE_H}
\begin{aligned}
    H &= -\sum_{\langle i,j\rangle,\sigma} t_{ij} 
    c_{i\sigma}^{\dagger} c_{j\sigma}
    - J \sum_{i,\sigma,\sigma'} 
    \mathbf{S}_i \cdot 
    c_{i\sigma}^{\dagger} 
    \bm{\sigma}_{\sigma\sigma'} 
    c_{i\sigma'} \\
    &\quad + (\epsilon - \mu) 
    \sum_{i,\sigma} 
    c_{i\sigma}^{\dagger} c_{i\sigma}.
\end{aligned}
\end{equation}
Here $c_{i\sigma}^{\dagger}$ ($c_{i\sigma}$) creates (annihilates) an electron with spin $\sigma=\uparrow,\downarrow$ at site $i$, defined with respect to a fixed spin-quantization axis, and $t_{ij}$ denotes the hopping amplitude between sites $i$ and $j$. 
For both materials, we include the intralayer nearest-neighbor hopping $t_1$ and the interlayer nearest-neighbor hopping $t_2$. 

The inequivalent geometric arrangement of nonmagnetic atoms between the magnetic sites carrying the same spin leads to bond-dependent hopping amplitudes, giving rise to the anisotropy responsible for altermagnetism. In Fe$_{1/4}$NbS$_2$, the leading anisotropic interlayer hoppings connect magnetic sites along two sets of symmetry-related bonds. For the $A$ sublattice, the bonds $\pm(3a/2,\sqrt{3}a/2,c)$, $\pm(-3a/2,\sqrt{3}a/2,c)$, and $\pm(0,-\sqrt{3}a,c)$ carry hopping amplitude $t_{3a}=t_3+\delta t$, whereas the bonds $\pm(3a/2,\sqrt{3}a/2,-c)$, $\pm(-3a/2,\sqrt{3}a/2,-c)$, and $\pm(0,-\sqrt{3}a,-c)$ carry $t_{3b}=t_3-\delta t$. For the $B$ sublattice, the assignment of $t_{3a}$ and $t_{3b}$ is reversed. In V$_{1/3}$NbS$_2$, the corresponding leading anisotropic interlayer hoppings 
occur along $\pm(a,0,\pm c)$, $\pm(a/2,\sqrt{3}a/2,\pm c)$, and $\pm(a/2,-\sqrt{3}a/2,\pm c)$. These anisotropic bonds are illustrated by the black and brown double arrows in Fig.~\ref{fig:lattice}. 

In addition, the parameter $J$ denotes the coupling between the itinerant electron spins and the localized magnetic moments $\mathbf{S}_i$, with $\bm{\sigma}$ representing the Pauli matrices. The localized moments follow the layered antiferromagnetic order described above. The on-site energy and chemical potential are denoted by $\epsilon$ and $\mu$, respectively, and all other hoppings are neglected for simplicity.

\subsection{$g$-Wave Spin Splitting}

We obtain the electronic band structure by diagonalizing the Hamiltonian matrix at each crystal momentum $\mathbf{k}$ (see the Appendices for details). In the absence of spin–orbit coupling, $\sigma$ remains a good quantum number and Eq.~(\ref{ELE_H}) reduces to
\begin{equation}
H=\sum_{n,\mathbf{k},\sigma}E_{n,\sigma}(\mathbf{k})f_{n,\mathbf{k},\sigma}^{\dagger}f_{n,\mathbf{k},\sigma},
\end{equation}
where $f_{n,\mathbf{k},\sigma}^{\dagger}~(f_{n,\mathbf{k},\sigma})$ is a fermionic creation (annihilation) operator and $E_{n,\sigma}(\mathbf{k})$ is the band dispersion. We take $t_1$ as the unit of energy.

\begin{figure}[htbp]
    \centering
    \includegraphics[width=1.0\linewidth]{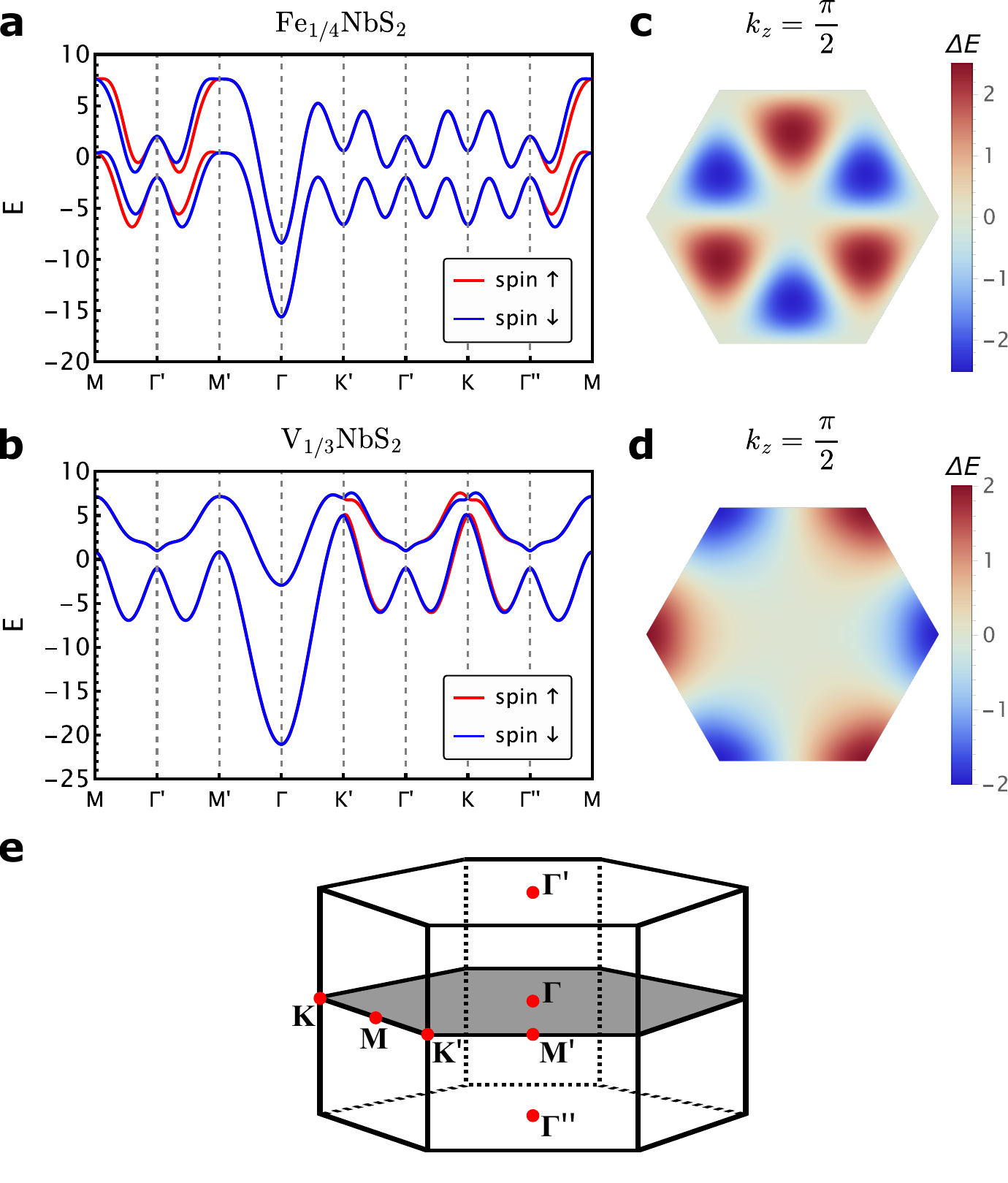}
    \caption{\textbf{Band dispersion of effective tight-binding models for Fe$_{1/4}$NbS$_2$ and V$_{1/3}$NbS$_2$.} \textbf{a} and \textbf{b} The electronic band structure along the chosen high-symmetry path in the first Brillouin zone. The parameters used are $t_2=1.5$, $t_3=0.5$, $\delta t=0.2$, $JS=1$, $\epsilon=0$, and $\mu=0$. The spin up and spin down bands are shown in red and blue, respectively. \textbf{c} and \textbf{d} False color plots of the spin-splitting energy $\Delta E_n(\mathbf{k})=E_{n,\uparrow}(\mathbf{k})-E_{n,\downarrow}(\mathbf{k})$ for the first pair of bands in the first Brillouin zone. \textbf{e} The first Brillion zone of Fe$_{1/4}$NbS$_2$ and V$_{1/3}$NbS$_2$.}
    \label{Fig:TB}
\end{figure}

The corresponding band structures of the effective Hamiltonian along the chosen high-symmetry path are presented in Figs.~\ref{Fig:TB}a and \ref{Fig:TB}b. For the Fe$_{1/4}$NbS$_2$ lattice, the spin splitting alternates along $\Gamma''$–$M$–$\Gamma'$–$M'$, whereas for the V$_{1/3}$NbS$_2$ lattice, the alternation occurs along $K'$–$\Gamma'$–$K$–$\Gamma''$, with these high symmetry points shown in Fig.~\ref{Fig:TB}e. The spin splitting of the lowest pair of bands, $\Delta E_1(\mathbf{k}) = E_{1\uparrow}(\mathbf{k}) - E_{1\downarrow}(\mathbf{k})$, evaluated on the $k_z=\pi/2$ plane of the first Brillouin zone, is shown in Figs.~\ref{Fig:TB}c and \ref{Fig:TB}d. In both cases, the spin splitting exhibits the characteristic $g$-wave sign-changing structure of altermagnetism.

However, the nodal-plane geometry differs between the two systems. For Fe$_{1/4}$NbS$_2$, the nodal planes are given by $k_y=0$, $k_y=\pm\sqrt{3}k_x$, and $k_z=0$, 
whereas for V$_{1/3}$NbS$_2$, they occur at 
$k_x=0$, $k_x=\pm\sqrt{3}k_y$, and $k_z=0$. 
This demonstrates that although both materials belong to the same $g$-wave altermagnetic class, the momentum-space nodal structure is material dependent within the class.

To further elucidate the microscopic origin of the splitting, we expand the Hamiltonian near the $\Gamma$ point. The leading-order spin splitting takes the form 
$\Delta E(\mathbf{k}) \propto \delta t\, k_y k_z (3k_x^2 - k_y^2)$ for Fe$_{1/4}$NbS$_2$ and 
$\Delta E(\mathbf{k}) \propto \delta t\, k_x k_z (3k_y^2 - k_x^2)$ for V$_{1/3}$NbS$_2$. These expressions explicitly show that the magnitude of the spin splitting 
is proportional to the anisotropic hopping difference $\delta t$, demonstrating that bond-dependent effective hopping plays a central role in generating altermagnetism.

\section{Low Energy Magnetic Excitations}

\begin{figure*}[t]
    \centering
    \includegraphics[width=1.0\linewidth]{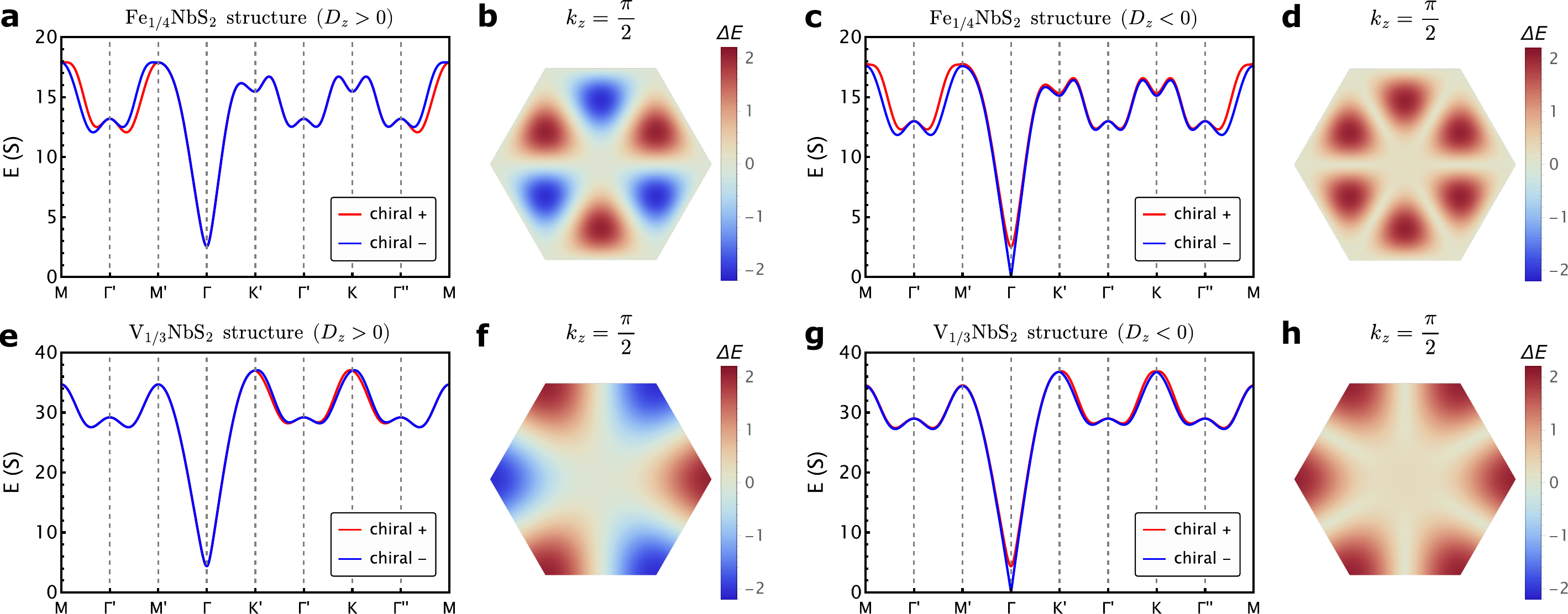}
    \caption{\textbf{Band dispersion of effective spin models for the Fe$_{1/4}$NbS$_2$ and V$_{1/3}$NbS$_2$ structures.} \textbf{a}, \textbf{c}, \textbf{e}, and \textbf{g} The magnon band structure along the chosen high-symmetry path in the first Brillouin zone. The parameters used are $J_1=-1.0$, $J_2=4.0$, $J_3=-0.2$, $\delta J=0.1$ and $|D_z|=0.2$. The chiral branches $+$ and $-$ are shown in red and blue, respectively. \textbf{b}, \textbf{d}, \textbf{f}, and \textbf{h} False color plots of the chiral-splitting energy $\Delta E(\mathbf{k})=E_{+}(\mathbf{k})-E_{-}(\mathbf{k})$ for the 1st pair of bands in the first Brillouin zone.}
    \label{Fig:LSW}
\end{figure*}

Having established the $g$-wave altermagnetic electronic structure within the effective tight-binding description, we now turn to the spin dynamics associated with the underlying magnetic order. In both materials, the altermagnetic state originates from ordered local moments on the transition-metal ions located between the NbS$_2$ layers. We therefore construct a Heisenberg Hamiltonian consistent with the magnetic structure and employ spin-wave theory to analyze the spin excitations,
\begin{equation}\label{HSPIN}
    H_{M}=\sum_{ ij}J_{ij}\bm{S}_i\cdot\bm{S}_j
    - D_z\sum_i(S_i^z)^2,
\end{equation}
where $J_{ij}$ represents the Heisenberg exchange interaction between sites $i$ and $j$, and $D_z$ denotes the strength of the single-ion anisotropy. Based on the magnetic ground states, we include the ferromagnetic intralayer nearest-neighbor exchange $J_1$ and the antiferromagnetic interlayer nearest-neighbor exchange $J_2$ for both materials. The anisotropic exchange $J_3\pm\delta J$ are defined on the same bonds as the anisotropic hopping terms in the tight-binding model, as illustrated in Fig.~\ref{fig:lattice}, reflecting the underlying bond-dependent anisotropy. Due to the out-of-plane (in-plane) spin orientation in the magnetic ground state of Fe$_{1/4}$NbS$_2$ (V$_{1/3}$NbS$_2$), we include a single-ion anisotropy term $D_z>0$ ($D_z<0$), respectively.

To investigate the collective spin excitations, we perform the Holstein--Primakoff transformation,
\begin{equation}
\begin{aligned}
    \tilde{S}_i^+&=(2S-a_i^{\dagger}a_i)^{\frac{1}{2}}a_i
    =\sqrt{2S}a_i-\frac{a_i^{\dagger}a_ia_i}{2\sqrt{2S}}
    +O(S^{-\frac{3}{2}}),\\
    \tilde{S}_i^-&=a_i^{\dagger}(2S-a_i^{\dagger}a_i)^{\frac{1}{2}}
    =\sqrt{2S}a_i^{\dagger}-\frac{a_i^{\dagger}a_i^{\dagger}a_i}{2\sqrt{2S}}
    +O(S^{-\frac{3}{2}}),\\
    \tilde{S}_i^z&=S-a_i^{\dagger}a_i,
\end{aligned}
\end{equation}
where $a_i^{\dagger}$ ($a_i$) is the magnon creation (annihilation) operator at site $i$, and the tilde indicates that the spin operators are expressed in the local coordinate frame aligned with the classical spin direction. The spin Hamiltonian in Eq.~(\ref{HSPIN}) can then be expanded in powers of $1/S$ as
\begin{equation}
    H_M = H_0 + H_1 + H_2 + \cdots,
\end{equation}
where $H_n$ denotes terms of order $S^{2-n}$. Here $H_0$ corresponds to the classical ground-state energy, $H_1$ describes the linear spin-wave Hamiltonian, and $H_2$ provides the leading $1/S$ correction to the magnon dispersion.

\subsection{Linear Spin-Wave Theory Analysis}

We first investigate the spin excitations within linear spin-wave theory (LSWT). 
After performing the Fourier transformation on the two magnetic sublattices with opposite spins separately, the quadratic spin-wave Hamiltonian $H_1$ can be written in momentum space as
\begin{equation}
    H_1=\sum_{\bm{k}} A^{\dagger}(\bm{k}) H(\bm{k}) A(\bm{k}),
\end{equation}
where $A(\bm{k})=(a_{\bm{k}},\, b_{\bm{k}},\, a_{-\bm{k}}^{\dagger},\, b_{-\bm{k}}^{\dagger})^{T}$, with $a_{\bm{k}}$ and $b_{\bm{k}}$ denoting the Fourier-transformed magnon operators on the two magnetic sublattices, and $H(\bm{k})$ is a $4\times4$ matrix. Then we diagonalize $H(\bm{k})$ through the Bogoliubov transformation, yielding
\begin{equation}
    H_1=E_{0}^{(1)}
    +\sum_{\bm{k}}\Big[E_{+}^{(0)}(\bm{k})\,\alpha_{\bm{k}}^{\dagger}\alpha_{\bm{k}}
    +E_{-}^{(0)}(\bm{k})\,\beta_{\bm{k}}^{\dagger}\beta_{\bm{k}}
    \Big],
\end{equation}
where $E_{0}^{(1)}$ is the order $S$ quantum correction to the classical ground state energy, and $E_{\pm}^{(0)}(\bm{k})$ denotes the magnon dispersions obtained within LSWT (see the Appendices for details). The two branches correspond to magnons with opposite chiralities, describing spin precessions with opposite rotational directions (counterclockwise and clockwise) relative to the classical spin direction on sublattice A, as determined by the Landau--Lifshitz equation. The magnon band dispersion in both material structures with different single ion anisotropy $D_z$ is shown in Fig.~\ref{Fig:LSW}.  

In the case of out-of-plane spin order ($D_z>0$), Figs.~\ref{Fig:LSW}a and \ref{Fig:LSW}e show the LSWT magnon band dispersions $E_{\pm}^{(0)}(\bm{k})$ along the chosen high-symmetry path for the Fe$_{1/4}$NbS$_2$ and V$_{1/3}$NbS$_2$ structures, respectively. 
For the Fe$_{1/4}$NbS$_2$ structure, the magnon chiral splitting alternates along $\Gamma''$–$M$–$\Gamma'$–$M'$, whereas for the V$_{1/3}$NbS$_2$ structure the alternation occurs along $K'$–$\Gamma'$–$K$–$\Gamma''$. 

Within LSWT, we obtain analytical expressions for the chiral splitting $\Delta E^{(0)}(\bm{k})=E_{+}^{(0)}(\bm{k})-E_{-}^{(0)}(\bm{k})$ for the two structures,
\begin{equation}
\begin{aligned}
    \Delta E_1^{(0)}(\bm{k})=&-16S\delta J(\cos(\frac{3}{2}k_x)-\cos(\frac{\sqrt{3}}{2}k_y))\\
    &\times\sin(\frac{\sqrt{3}}{2}k_y)\sin(k_z),
\end{aligned}
\end{equation}
\begin{equation}
\begin{aligned}
    \Delta E_2^{(0)}(\bm{k})=&-16S\delta J(\cos(\frac{1}{2}k_x)-\cos(\frac{\sqrt{3}}{2}k_y))\\
    &\times\sin(\frac{1}{2}k_x)\sin(k_z),
\end{aligned}
\end{equation}
where the indices $1$ and $2$ label the Fe$_{1/4}$NbS$_2$ and V$_{1/3}$NbS$_2$ structures, respectively. 
The chiral splitting evaluated on the $k_z=\pi/2$ plane is shown in Figs.~\ref{Fig:LSW}b and \ref{Fig:LSW}f. 

Similar to the electronic bands, the magnon chiral splitting exhibits the characteristic $g$-wave sign-changing structure. The nodal-plane structure of the magnon chiral splitting coincides with that of the electronic $g$-wave altermagnetic splitting discussed above. In particular, if $D_z\ge0$, the LSWT chiral splitting $\Delta E^{(0)}(\bm{k})$ depends only on the anisotropic exchange difference $\delta J$ between bonds connecting the same sublattices, and is independent of the other exchange parameters and the single-ion anisotropy $D_z$.

In the case of easy-plane sing-ion anisotropy ($D_z<0$), the magnon dispersions along the chosen high-symmetry path and the corresponding magnon splitting on the $k_z=\pi/2$ plane are shown in Figs.~\ref{Fig:LSW}c, \ref{Fig:LSW}g, \ref{Fig:LSW}d, and \ref{Fig:LSW}h. In contrast to the $D_z>0$ case, the chiral splitting satisfies $\Delta E^{(0)}(\bm{k}) \ge 0$ throughout the Brillouin zone. Consequently, the in-plane spin order does not produce the characteristic $g$-wave splitting structure, and no nodal planes appear inside the first Brillouin zone. This distinction arises from differences in transverse spin fluctuations between the easy-axis and easy-plane states, as discussed further in the Discussion section.

\subsection{$1/S$ Correction from Magnon-Magnon Interactions to $g$-Wave Chiral Splitting}

While LSWT captures the characteristic $g$-wave chiral splitting of the magnon bands when $D_z\ge0$, it neglects the effects of magnon-magnon interactions. To examine the robustness of the chiral splitting for the out-of-plane spin order against magnon-magnon interactions, we now go beyond LSWT and consider the leading $1/S$ correction arising from the quartic terms $H_2$ in the spin-wave Hamiltonian. In particular, we focus on how these interactions renormalize the magnon dispersion and modify the $g$-wave splitting between the two chiral branches.

For the Fe$_{1/4}$NbS$_2$ structure with $D_z\ge0$, the Bogoliubov transformation that diagonalizes the LSWT Hamiltonian can be written as
\begin{equation}\label{BG}
a_{\bm{k}}^{\dagger}
=
u_{\bm{k}}\alpha_{\bm{k}}^{\dagger}
+
v_{\bm{k}}\beta_{-\bm{k}},\qquad
b_{-\bm{k}}
=
v_{\bm{k}}\alpha_{\bm{k}}^{\dagger}
+
u_{\bm{k}}\beta_{-\bm{k}},
\end{equation}
with
\begin{equation}
u_{\bm{k}}
=
\left(\frac{1+\epsilon_{\bm{k}}}{2\epsilon_{\bm{k}}}\right)^{1/2},
\qquad
v_{\bm{k}}
=
-\mathrm{sgn}(\gamma_{\bm{k}})
\left(\frac{1-\epsilon_{\bm{k}}}{2\epsilon_{\bm{k}}}\right)^{1/2},
\end{equation}
where
\begin{equation}
\epsilon_{\bm{k}}
=
(1-\gamma_{\bm{k}}^2)^{1/2},
\qquad
\gamma_{\bm{k}}
=\frac{g_{\bm{k}}}{f_{\bm{k}}}.
\end{equation}
Here $\alpha$ and $\beta$ denote the magnon eigenmodes with chirality $\pm$, while $f_{\bm{k}}$ and $g_{\bm{k}}$ are matrix elements of the LSWT Hamiltonian $H_1(\bm{k})$ (see the Appendices for details).

To evaluate the leading $1/S$ correction to the magnon energies, we substitute the Bogoliubov transformation in Eq.~(\ref{BG}) into the quartic magnon interaction term $H_2$ and bring the resulting expression into normal order,
\begin{equation}
H_2=\text{const}~+:H_1':+:H_2':,
\end{equation}
where
\begin{equation}
:H_1':=
\sum_{\bm{k}}A_{\bm{k}}^+\alpha_{\bm{k}}^{\dagger}\alpha_{\bm{k}}+A_{\bm{k}}^-\beta_{\bm{k}}^{\dagger}\beta_{\bm{k}}+B_{\bm{k}}(\alpha_{\bm{k}}^{\dagger}\beta_{-\bm{k}}^{\dagger}+\alpha_{\bm{k}}\beta_{-\bm{k}}).
\end{equation}
This procedure generates additional quadratic terms in the magnon eigenmodes. The term $:H_1':$ is known as the Oguchi correction, where $A_{\bm{k}}^{\pm}$ provide the leading $1/S$ renormalization of the magnon dispersion. The remaining term $:H_2':$ describes the residual quartic magnon--magnon interactions. In the collinear magnetic system, $A_{\bm{k}}^{\pm}$ can be decomposed into three parts,
\begin{equation}
A_{\bm{k}}^{\pm}=\sum_{\bm{\Delta}}A_{\bm{k},\mathrm{P}}^{\pm}(\bm{\Delta})+\sum_{\bm{\delta}}A_{\bm{k},\mathrm{AP}}^{\pm}(\bm{\delta})+A_{\bm{k},D}^{\pm},
\end{equation}
where $A_{\bm{k},\mathrm{P}}^{\pm}$ ($A_{\bm{k},\mathrm{AP}}^{\pm}$) arises from the Heisenberg term $\bm{S}_i\cdot\bm{S}_j$ at the bond $\bm{\Delta}$ ($\bm{\delta}$) between the sites 
with parallel (antiparallel) magnetic moments, and $A_{\bm{k},D}^{\pm}$ is the contribution from the single-ion anisotropy term. They are given by

\begin{equation}
\label{eq:fm}
\begin{aligned}
    A_{\bm{k},\mathrm{P}}^+(\Delta)&=(J_{A\bm{\Delta}}\frac{1+\epsilon_{\bm{k}}}{2\epsilon_{\bm{k}}}+J_{B\bm{\Delta}}\frac{1-\epsilon_{\bm{k}}}{2\epsilon_{\bm{k}}})(1-\cos(\bm{k}\bm{\Delta}))\\
    &\times\frac{1}{N}\sum_{\bm{q}}\frac{(1-\epsilon_{\bm{q}})(1-\cos(\bm{q}\bm{\Delta}))}{\epsilon_{\bm{q}}},\\
    A_{\bm{k},\mathrm{P}}^-(\Delta)&=(J_{B\bm{\Delta}}\frac{1+\epsilon_{\bm{k}}}{2\epsilon_{\bm{k}}}+J_{A\bm{\Delta}}\frac{1-\epsilon_{\bm{k}}}{2\epsilon_{\bm{k}}})
    (1-\cos(\bm{k}\bm{\Delta}))\\
    &\times\frac{1}{N}\sum_{\bm{q}}\frac{(1-\epsilon_{\bm{q}})(1-\cos(\bm{q}\bm{\Delta}))}{\epsilon_{\bm{q}}},
\end{aligned}
\end{equation}

\begin{equation}
\label{eq:afm}
A_{\bm{k},\mathrm{AP}}^{\pm}(\bm{\delta})=J_{\bm{\delta}}\frac{1-\gamma_{\bm{k}}\cos(\bm{k}\bm{\delta})}{2\epsilon_{\bm{k}}}\frac{1}{N}\sum_{\bm{q}}\frac{\gamma_{\bm{q}}\cos(\bm{q}\bm{\delta})+\epsilon_{\bm{q}}-1}{\epsilon_{\bm{q}}},
\end{equation}

\begin{equation}
\label{eq:d}
    A_{\bm{k},D}^{\pm}=-\frac{D_z}{\epsilon_{\bm{k}}}(1+\frac{2}{N}\sum_{\bm{q}}\frac{1-\epsilon_{\bm{q}}}{\epsilon_{\bm{q}}}),
\end{equation}
where $J_{A\bm{\Delta}}$ ($J_{B\bm{\Delta}}$) are the Heisenberg exchange at bond $\bm{\Delta}$ on A (B) sublattices.

\begin{figure}[htbp]
    \centering
    \includegraphics[width=1.0\linewidth]{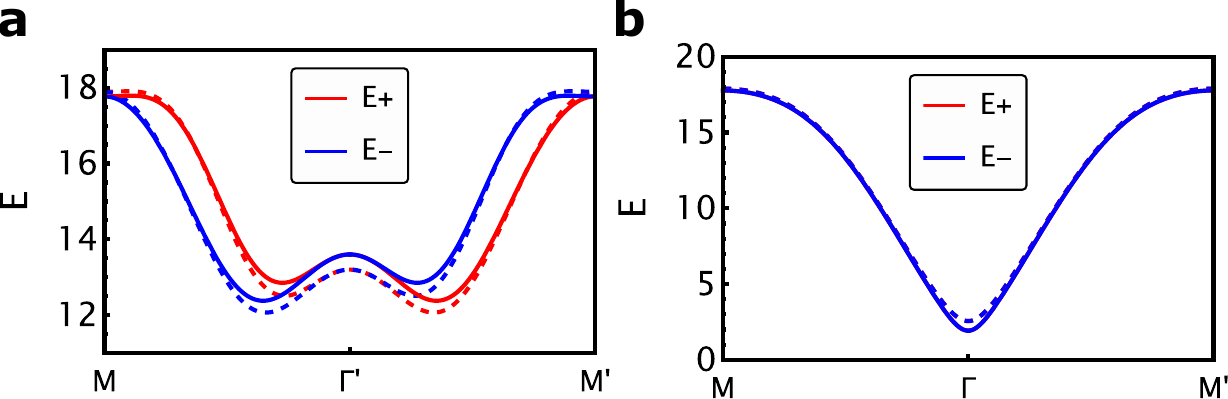}
    \caption{\textbf{Renormalized magnon band dispersion of the effective spin model for the Fe$_{1/4}$NbS$_2$ structure after $1/S$ correction.} \textbf{a} Magnon band structures along the $\mathbf{k}$ path M--$\Gamma'$--M$'$. \textbf{b} Magnon band structures along the $\mathbf{k}$ path M--$\Gamma$--M$'$. The parameters are $J_1=-1.0$, $J_2=4.0$, $J_3=-0.2$, $\delta J=0.1$, $D_z=0.2$ and $S=1$. The chiral branches $+$ and $-$ are shown in red and blue, respectively, while the bands with and without $1/S$ correction are represented by solid and dashed lines.}
    \label{Fig:corr1}
\end{figure}

Fig.~\ref{Fig:corr1} compares the renormalized magnon band structures with LSWT along two representative momentum paths. The path in Fig.~\ref{Fig:corr1}$\bm{a}$ is chosen such that the LSWT spectrum exhibits chiral splitting, whereas the path in Fig.~\ref{Fig:corr1}$\bm{b}$ corresponds to a direction where the bands are degenerate. Including the $1/S$ correction leads to a quantitative renormalization of the dispersion while preserving these key features, with the splitting in Fig.~\ref{Fig:corr1}$\bm{a}$ remaining clearly visible and the degeneracy in Fig.~\ref{Fig:corr1}$\bm{b}$ maintained. This indicates that the $1/S$ correction primarily modifies the energy scale without altering the symmetry-determined structure of the magnon spectrum.

From Eqs.~(\ref{eq:fm})-(\ref{eq:d}), only the $1/S$ correction from spin-parallel bonds contributes to the renormalization of the chiral splitting, which is given by
\begin{equation}
\begin{aligned}
    \Delta E_1(\bm{k}) =&\, \Delta E_1^{(0)}(\bm{k}) +8 \delta J\, C_1 \left[\cos\left(\tfrac{3}{2}k_x\right) - \cos\left(\tfrac{\sqrt{3}}{2}k_y\right)\right] \\
    &\times \sin\left(\tfrac{\sqrt{3}}{2}k_y\right)\sin(k_z),
\end{aligned}
\end{equation}
where
\begin{equation}
\begin{aligned}
    C_1=&\frac{1}{N}\sum_{\bm{q}}\frac{1 - \epsilon_{\bm{q}}}{\epsilon_{\bm{q}}}\times\\
    &\left[1-\frac{1}{3}(2\cos\frac{3}{2}k_x\cos\frac{\sqrt{3}}{2}k_y+\cos\sqrt{3}k_y)\cos k_z\right]
\end{aligned}
\end{equation}
is a constant that depends on the model parameters. Similarly, for the V$_{1/3}$NbS$_2$ structure with $D_z\geq0$,
\begin{equation}
\begin{aligned}
    \Delta E_2(\bm{k}) =&\, \Delta E_2^{(0)}(\bm{k})+8 \delta J\, C_2 \left[\cos\left(\tfrac{1}{2}k_x\right) - \cos\left(\tfrac{\sqrt{3}}{2}k_y\right)\right] \\
    &\times \sin\left(\tfrac{1}{2}k_x\right)\sin(k_z),
\end{aligned}
\end{equation}
\begin{equation}
\begin{aligned}
    C_2=&\frac{1}{N}\sum_{\bm{q}}\frac{1 - \epsilon_{\bm{q}}}{\epsilon_{\bm{q}}}\times\\
    &\left[1-\frac{1}{3}(2\cos\frac{1}{2}k_x\cos\frac{\sqrt{3}}{2}k_y+\cos k_x)\cos k_z\right].
\end{aligned}
\end{equation}
Unlike in LSWT where the chiral splitting $\Delta E_i^{(0)}$ depends solely on the bond anisotropy $\delta J$, the inclusion of magnon-magnon interactions introduces a dependence on all model parameters. 

\begin{figure}[htbp]
    \centering
    \includegraphics[width=1.0\linewidth]{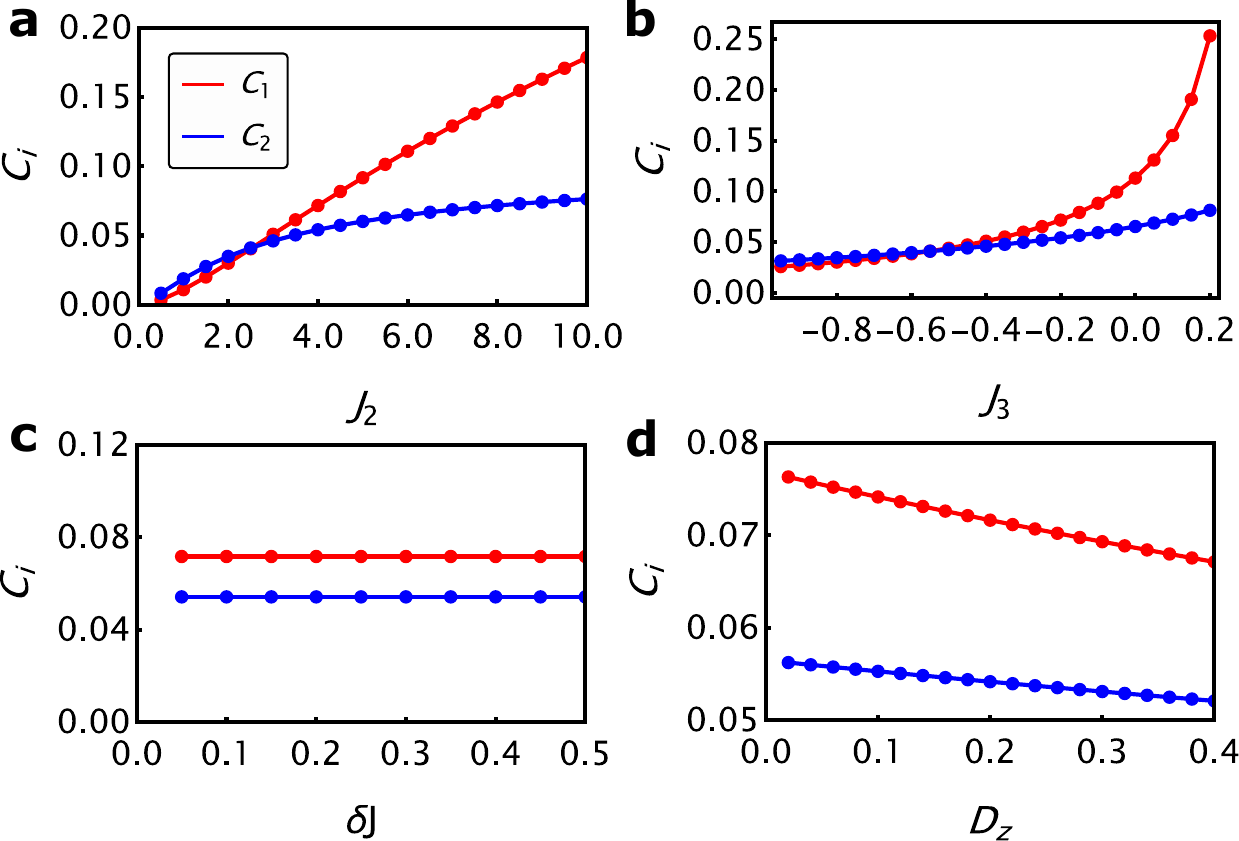}
    \caption{
\textbf{Dependence of the $1/S$ correction coefficients $C_i$ for the magnon chiral splitting on effective spin model parameters in the Fe$_{1/4}$NbS$_2$ and V$_{1/3}$NbS$_2$ structures.}
(a) Dependence on $J_2$, (b) dependence on $J_3$, (c) dependence on $\delta J$, and (d) dependence on $D_z$. The remaining parameters are fixed to $J_1=-1.0$, $J_2=4.0$, $J_3=-0.2$, $\delta J=0.1$, $D_z=0.2$, and $S=1$. The coefficients $C_1$ and $C_2$, corresponding to the Fe$_{1/4}$NbS$_2$ and V$_{1/3}$NbS$_2$ structures, are shown in red and blue, respectively.
}
    \label{Fig:corr2}
\end{figure}

The relative change of the chiral splitting is given by
\begin{equation}
    \frac{\Delta E_i(\bm{k})-\Delta E_i^{(0)}(\bm{k})}{\Delta E_i^{(0)}(\bm{k})} = -\frac{C_i}{2S},
\end{equation}
and its parameter dependence for both material structures is shown in Figure~\ref{Fig:corr2}. The $1/S$ correction leads to a reduction of the chiral splitting, indicating that magnon-magnon interaction suppresses the splitting amplitude. 

Quantitatively, $C_i$ increases significantly with increasing antiferromagnetic exchange $J_2$ and with increasing $J_3$ in the antiferromagnetic regime ($J_3>0$). In contrast, in the ferromagnetic regime ($J_3<0$), $C_i$ is only weakly enhanced as the magnitude $|J_3|$ decreases. A similar weak increase is observed upon reducing the single-ion anisotropy $D_z$. On the other hand, $C_i$ shows negligible dependence on the anisotropy $\delta J$, indicating that the relative correction is largely insensitive to the microscopic origin of the chiral splitting.

These results demonstrate that the magnon chiral splitting remains robust in the presence of magnon--magnon interactions, but its magnitude is systematically reduced, particularly in regimes dominated by antiferromagnetic exchange, where the $1/S$ correction becomes non-negligible.

\section{First-Principles Calculations}

To verify that the symmetry-based effective model captures the essential electronic features of the real materials, we further perform first-principles density functional theory (DFT) calculations for Fe$_{1/4}$NbS$_2$ and V$_{1/3}$NbS$_2$ using the Vienna \textit{ab initio} Simulation Package (VASP) with projector-augmented wave (PAW) potentials~\cite{PAW} and the revised Perdew--Burke--Ernzerhof (PBE) functional~\cite{PBE}. An on-site Coulomb repulsion of 4.0~eV was applied to the Fe and V $d$ orbitals according to Dudarev's rotationally invariant approach~\cite{LDAU}. Van der Waals interactions between NbS$_2$ layers were included using the DFT-D3 method of Grimme \textit{et al.} with Becke--Johnson damping~\cite{vdW1, vdW2}. 

All calculations were performed within a spin-polarized framework under the antiferromagnetic magnetic configuration, without including spin orbit coupling (SOC). A $15\times15\times6$ Monkhorst--Pack $k$-point mesh and a plane-wave energy cutoff of 450~eV were employed, and structural relaxation was carried out until the residual forces were smaller than $1$~meV/\AA. The fully relaxed lattice constants are $a=b=6.73$~\AA, $c=12.12$~\AA\ for Fe$_{1/4}$NbS$_2$, and $a=b=5.73$~\AA, $c=12.14$~\AA\ for V$_{1/3}$NbS$_2$, with $\alpha=\beta=90^\circ$ and $\gamma=120^\circ$ in both cases. These structural parameters are consistent with experimentally reported values~\cite{V_structure, Fe_structure}, supporting the reliability of the chosen exchange--correlation functional and Hubbard $U$.

\begin{figure}
    \centering
    \includegraphics[width=0.8\linewidth]{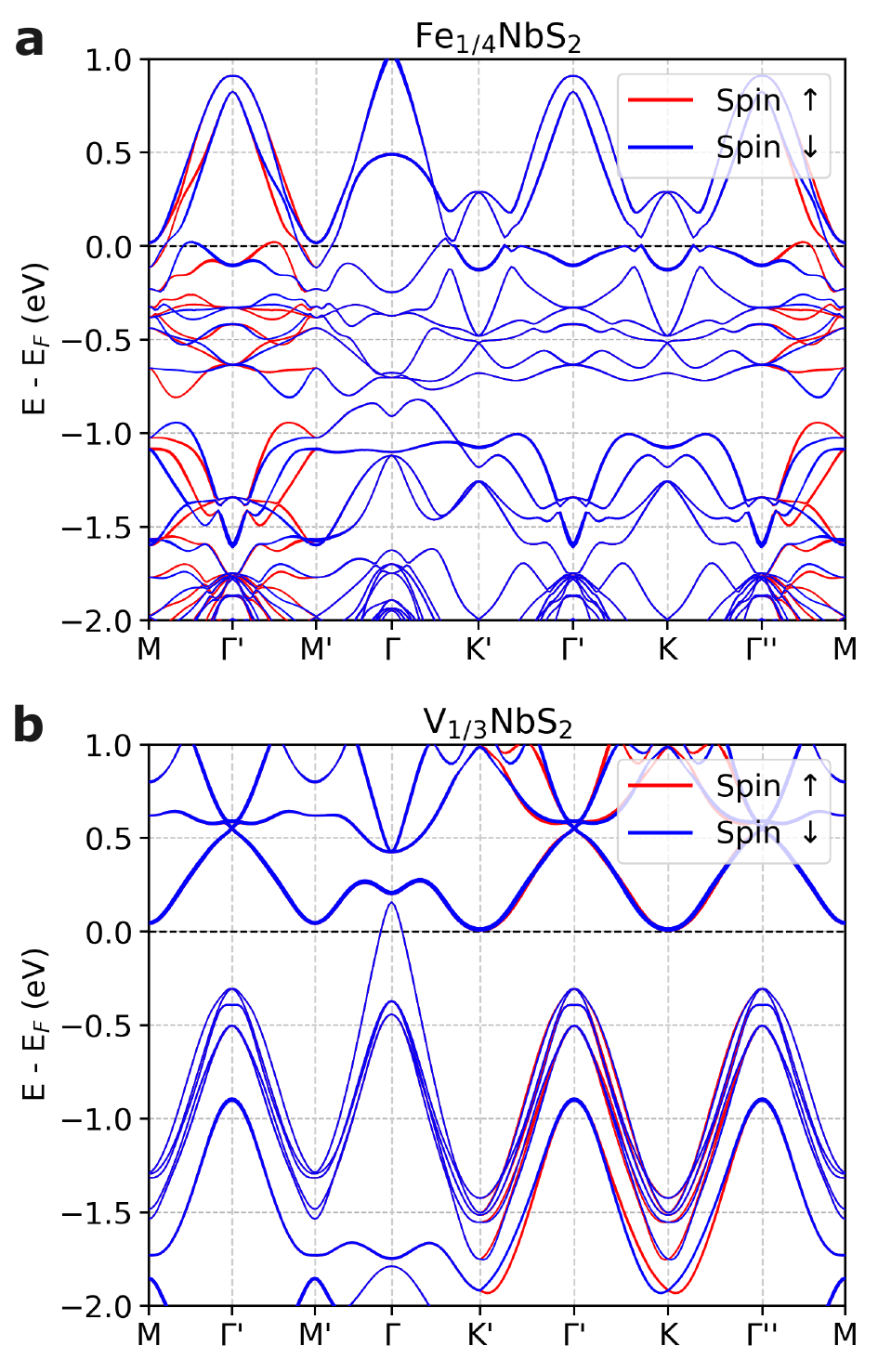}
    \caption{\textbf{First-principles band structures of Fe$_{1/4}$NbS$_2$ and V$_{1/3}$NbS$_2$.} \textbf{a} and \textbf{b} Spin-polarized electronic band structures calculated from DFT along the chosen high-symmetry path in the first Brillouin zone under the antiferromagnetic configuration without spin--orbit coupling, for Fe$_{1/4}$NbS$_2$ and V$_{1/3}$NbS$_2$, respectively. The spin-up and spin-down bands are shown in red and blue, respectively.}
    \label{fig:2}
\end{figure}

The resulting spin-polarized band structures of Fe$_{1/4}$NbS$_2$ and V$_{1/3}$NbS$_2$ along the selected high-symmetry path are shown in Fig.~\ref{fig:2} \textbf{a} and \textbf{b}, respectively. In both materials, a clear spin splitting of the electronic bands is observed despite the absence of net magnetization, which is a defining feature of altermagnetism. 

For Fe$_{1/4}$NbS$_2$, the spin splitting alternates in sign 
along the $\Gamma''$–$M$–$\Gamma'$–$M'$ path, forming a characteristic $g$-wave pattern. A similar sign-changing behavior is present in V$_{1/3}$NbS$_2$, although the momentum directions along which the splitting vanishes are different, reflecting the distinct underlying lattice geometries. 

Overall, the first-principles band structures exhibit the same 
$g$-wave symmetry and material-dependent nodal characteristics as those obtained from the effective tight-binding model, providing independent support for the bond-anisotropy mechanism proposed above.

\section{Discussion}

In this work we investigated the altermagnetic electronic and magnonic properties of Fe$_{1/4}$NbS$_2$ and V$_{1/3}$NbS$_2$ based on effective tight-binding and spin models. The tight-binding analysis shows that both systems host $g$-wave spin splitting originating from bond-dependent hopping anisotropy, while the corresponding nodal-plane structures are material dependent due to their distinct lattice geometries.

On the magnon side, LSWT reveals that a $g$-wave chiral splitting emerges for out-of-plane spin configurations ($D_z \ge 0$), with nodal structures directly inherited from the electronic counterpart. In contrast, for in-plane spin configurations ($D_z < 0$), the splitting becomes non-negative throughout the Brillouin zone and the characteristic $g$-wave structure disappears. 

This difference can be understood from the nature of spin fluctuations induced by single-ion anisotropy. In the easy-axis case, the magnetic moments align along the $z$ direction, and the two in-plane components of the transverse spin fluctuations are equivalent. As a result, the magnon modes remain degenerate in the conventional antiferromagnetic limit ($\delta J=0$), and develop a nontrivial splitting once the altermagnetic anisotropy is introduced. In contrast, for easy-plane anisotropy, the spins lie in the plane, leading to inequivalent transverse fluctuation channels. Even in the conventional antiferromagnetic case, this results in one Goldstone mode and one gapped mode. The inequivalence between these fluctuation modes modifies the magnon structure and destroys the well-defined altermagnetic chiral splitting. This analysis highlights that the altermagnetic nodal structure of the magnon bands is governed not only by symmetry considerations but also by the structure of spin fluctuations determined by magnetic anisotropy. A related behavior for in-plane spin configurations has been discussed in Ref.~\cite{vb_am1}, where the magnetic moment of each magnon branch becomes momentum dependent and exhibits a $g$-wave-like distribution.

Going beyond LSWT, we included magnon--magnon interactions at the $1/S$ level and found that the chiral splitting remains robust, with its momentum-dependent structure unchanged. However, the magnitude of the splitting is systematically reduced by quantum fluctuations. Importantly, while the LSWT splitting depends solely on the anisotropic exchange $\delta J$, the $1/S$ correction introduces dependence on additional parameters. In particular, the correction is strongly enhanced by increasing the antiferromagnetic exchange $J_2$ and $J_3$, while it shows weaker dependence on ferromagnetic $J_3$ and single-ion anisotropy $D_z$, and remains insensitive to $\delta J$. 

These results indicate that altermagnetic magnon splitting is a robust feature protected by symmetry, but its magnitude can be significantly renormalized in regimes dominated by antiferromagnetic interactions. In particular, for systems with relatively small local moments, where $1/S$ corrections are more pronounced, such interaction effects become important and should be taken into account when comparing with experimental measurements such as neutron scattering or resonant inelastic X-ray scattering.

Finally, our first-principles calculations provide further confirmation that the $g$-wave altermagnetic features identified in the effective model persist in realistic materials. This consistency indicates that the bond-anisotropy mechanism captures the essential low-energy physics and provides a reliable framework for describing altermagnetism in these systems.

\section{Acknowledgments}

This work is supported by the Center for Molecular Magnetic Quantum Materials, an Energy Frontier Research Center funded by the U.S. Department of Energy, Office of Science, Basic Energy Sciences under Award no. DE-SC0019330. Computations were done using the utilities of the University of Florida Research Computing.

\appendix
\renewcommand{\theequation}{A\arabic{equation}}
\setcounter{equation}{0}

\section{Tight Binding Model}
For both Fe$_{1/4}$NbS$_2$ and V$_{1/3}$NbS$_2$, the effective tight-binding Hamiltonian is
\begin{equation}\label{app:ELE_H}
\begin{aligned}
    H &= -\sum_{\langle i,j\rangle,\sigma} t_{ij} 
    c_{i\sigma}^{\dagger} c_{j\sigma}
    - J \sum_{i,\sigma,\sigma'} 
    \mathbf{S}_i \cdot 
    c_{i\sigma}^{\dagger} 
    \bm{\sigma}_{\sigma\sigma'} 
    c_{i\sigma'} \\
    &\quad + (\epsilon - \mu) 
    \sum_{i,\sigma} 
    c_{i\sigma}^{\dagger} c_{i\sigma},
\end{aligned}
\end{equation}
where the magnetic unit-cell vectors are chosen as $\mathbf{a}_1=(a,0,0)$, $\mathbf{a}_2=(-a/2,\sqrt{3}a/2,0)$, and $\mathbf{a}_3=(0,0,c)$. In the following calculations, we set $a=1$. We perform a Fourier transformation of the electron operators,
\begin{equation}\label{Fourier}
    c_{i\sigma}=\frac{1}{\sqrt{N}}\sum_{\mathbf{k}}c_{\mathbf{k}\sigma}e^{i\mathbf{k}\cdot\mathbf{r}_i},
\end{equation}
where $N$ is the number of unit cells. The Hamiltonian in Eq.~(\ref{app:ELE_H}) then becomes
\begin{equation}\label{app:Hk}
H=\sum_{\mathbf{k},\sigma}\psi_{\mathbf{k}\sigma}^{\dagger}H_{\mathbf{k}\sigma}\psi_{\mathbf{k}\sigma},
\end{equation}
where $\psi_{\mathbf{k}\sigma}=(c_{1\sigma\mathbf{k}}, c_{2\sigma\mathbf{k}})^{T}$. The matrices $H_{\mathbf{k}\sigma}$ are given by
\begin{equation}
    H_{\mathbf{k},\sigma}=
    \begin{bmatrix}
        f_{1\bm{k}} + f_{3a\bm{k}} & f_{2\bm{k}} \\
        f_{2\bm{k}}^* & f_{1\bm{k}} + f_{3b\bm{k}} \\
        \end{bmatrix}\mp JS\tau_3+(\epsilon-\mu)\tau_0,
\end{equation}
where
\begin{equation}
\begin{aligned}
f_{1\bm{k}}&=-2t_1\sum_{i}\cos(\bm{k}\bm{d}_i),\\
f_{2\bm{k}}&=-t_2\sum_{i}\exp(-i\bm{k}\bm{\delta}_i),\\
f_{3a\bm{k}}&=-2(t_3+\delta t)\sum_{i}\cos(\bm{k}\bm{\Delta}_{Ai})\\
&\quad-2(t_3-\delta t)\sum_{i}\cos(\bm{k}\bm{\Delta}_{Bi}),\\
f_{3b\bm{k}}&=-2(t_3-\delta t)\sum_{i}\cos(\bm{k}\bm{\Delta}_{Ai})\\
&\quad-2(t_3+\delta t)\sum_{i}\cos(\bm{k}\bm{\Delta}_{Bi}).\\
\end{aligned}
\end{equation}
The bond vectors are given by
\begin{equation}
   \bm{d}_1=(1,0,0),~\bm{d}_2=(\tfrac{1}{2},\tfrac{\sqrt{3}}{2},0),~\bm{d}_3=(\tfrac{1}{2},-\tfrac{\sqrt{3}}{2},0),
\end{equation}

for Fe$_{1/4}$NbS$_2$, 
\begin{equation}
   \bm{\delta}_1=(0,0,\tfrac{1}{2}),~\bm{\delta}_2=(0,0,-\tfrac{1}{2}), 
\end{equation}
\begin{equation}
\begin{aligned}
   \bm{\Delta}_{A1}&=(\tfrac{3}{2},\tfrac{\sqrt{3}}{2},1),~
   \bm{\Delta}_{A2}=(-\tfrac{3}{2},\tfrac{\sqrt{3}}{2},1),\\ 
   \bm{\Delta}_{A3}&=(0,-\sqrt{3},1),~
   \bm{\Delta}_{B1}=(\tfrac{3}{2},\tfrac{\sqrt{3}}{2},-1),\\
   \bm{\Delta}_{B2}&=(-\tfrac{3}{2},\tfrac{\sqrt{3}}{2},-1),~
   \bm{\Delta}_{B3}=(0,-\sqrt{3},-1),
\end{aligned}
\end{equation}

and for V$_{1/3}$NbS$_2$, 
\begin{equation}
\begin{aligned}
    \bm{\delta}_1&=(\tfrac{1}{2},\tfrac{1}{2\sqrt{3}},\tfrac{1}{2}),~
    \bm{\delta}_2=(-\tfrac{1}{2},\tfrac{1}{2\sqrt{3}},\tfrac{1}{2}),\\
    \bm{\delta}_3&=(0,-\tfrac{1}{\sqrt{3}},\tfrac{1}{2}),~
    \bm{\delta}_4=(\tfrac{1}{2},\tfrac{1}{2\sqrt{3}},-\tfrac{1}{2}),\\
    \bm{\delta}_5&=(-\tfrac{1}{2},\tfrac{1}{2\sqrt{3}},-\tfrac{1}{2}),~
    \bm{\delta}_6=(0,-\tfrac{1}{\sqrt{3}},-\tfrac{1}{2}),
\end{aligned}
\end{equation}

\begin{equation}
\begin{aligned}
   \bm{\Delta}_{A1}&=(1,0,1),~
   \bm{\Delta}_{A2}=(-\tfrac{1}{2},\tfrac{\sqrt{3}}{2},1),\\ 
   \bm{\Delta}_{A3}&=(-\tfrac{1}{2},-\tfrac{\sqrt{3}}{2},1),~
   \bm{\Delta}_{B1}=(1,0,-1),\\
   \bm{\Delta}_{B2}&=(-\tfrac{1}{2},\tfrac{\sqrt{3}}{2},-1),~
   \bm{\Delta}_{B3}=(-\tfrac{1}{2},-\tfrac{\sqrt{3}}{2},-1).
\end{aligned}
\end{equation}
After diagonalization, the Hamiltonian becomes
\begin{equation}
H=\sum_{n,\mathbf{k},\sigma}E_{n,\sigma}(\mathbf{k})f_{n,\mathbf{k},\sigma}^{\dagger}f_{n,\mathbf{k},\sigma},
\end{equation}
and the band dispersion is given by
\begin{equation}
\begin{aligned}
    E_{1,\sigma}(\bm{k})&=f_{1\bm{k}}+f_{3\bm{k}}-\sqrt{(\delta f_{3\bm{k}}\mp JS)^2+|f_{2\bm{k}}|^2},\\
    E_{2,\sigma}(\bm{k})&=f_{1\bm{k}}+f_{3\bm{k}}+\sqrt{(\delta f_{3\bm{k}}\mp JS)^2+|f_{2\bm{k}}|^2},
\end{aligned}
\end{equation}
where $f_{3\bm{k}}=(f_{3a\bm{k}}+f_{3b\bm{k}})/2$ and $\delta f_{3\bm{k}}=(f_{3a\bm{k}}-f_{3b\bm{k}})/2$.

\section{Linear Spin-Wave Theory}

For both Fe$_{1/4}$NbS$_2$ and V$_{1/3}$NbS$_2$ structures, the effective Heisenberg Hamiltonian is given by
\begin{equation}\label{app:HSPIN}
    H_{M}=\sum_{\langle ij\rangle}J_{ij}\bm{S}_i\cdot\bm{S}_j
    - D_z\sum_i(S_i^z)^2.
\end{equation}
After performing the Holstein--Primakoff transformation, we obtain the linear spin-wave Hamiltonian $H_1$. We then apply a Fourier transformation to the bosonic operators,
\begin{equation}
   a_{i}=\frac{1}{\sqrt{N}}\sum_{\mathbf{k}}a_{\mathbf{k}}e^{i\mathbf{k}\cdot\mathbf{r}_i},\quad
   b_{j}=\frac{1}{\sqrt{N}}\sum_{\mathbf{k}}b_{\mathbf{k}}e^{i\mathbf{k}\cdot\mathbf{r}_j},
\end{equation}
and $H_1$ can be written in momentum space as
\begin{equation}
    H_1=\sum_{\bm{k}} A^{\dagger}(\bm{k}) H(\bm{k}) A(\bm{k}),
\end{equation}
where $A(\bm{k})=(a_{\bm{k}},\, b_{\bm{k}},\, a_{-\bm{k}}^{\dagger},\, b_{-\bm{k}}^{\dagger})^{T}$, and $H(\bm{k})$ is a $4\times4$ matrix,
\begin{equation}
    H^{\text{op}}(\bm{k})=
    \begin{bmatrix}
        g_{a\bm{k}} & 0 & 0 & g_{2\bm{k}}\\
        0 & g_{b\bm{k}} & g_{2-\bm{k}} & 0\\
        0 & g_{2-\bm{k}}^* & g_{a-\bm{k}} & 0\\
        g_{2\bm{k}}^* & 0 & 0 & g_{b-\bm{k}}\\
        \end{bmatrix}+2SD_z\mathbbm{1}_{4\times4},
\end{equation}

\begin{equation}
    H^{\text{ip}}(\bm{k})=
    \begin{bmatrix}
        g_{a\bm{k}} & 0 & -SD_z & g_{2\bm{k}}\\
        0 & g_{b\bm{k}} & g_{2-\bm{k}} & -SD_z\\
        -SD_z & g_{2-\bm{k}}^* & g_{a-\bm{k}} & 0\\
        g_{2\bm{k}}^* & -SD_z & 0 & g_{b-\bm{k}}\\
        \end{bmatrix}-SD_z\mathbbm{1}_{4\times4},
\end{equation}
where ``op'' and ``ip'' denote the out-of-plane and in-plane spin configurations, respectively, and $\mathbbm{1}_{4\times4}$ is the $4\times4$ identity matrix. The matrix elements are given for the Fe$_{1/4}$NbS$_2$ structure by
\begin{equation}
\begin{aligned}
   g_{a\bm{k}}=&-6SJ_1+2SJ_2-12SJ_3+2SJ_1\sum_i\cos(\bm{k}\bm{d}_i)\\
   &+2S(J_3+\delta J)\sum_i\cos(\bm{k}\bm{\Delta}_{Ai})\\
   &+2S(J_3-\delta J)\sum_i\cos(\bm{k}\bm{\Delta}_{Bi}),\\
   g_{b\bm{k}}=&-6SJ_1+2SJ_2-12SJ_3+2SJ_1\sum_i\cos(\bm{k}\bm{d}_i)\\
   &+2S(J_3-\delta J)\sum_i\cos(\bm{k}\bm{\Delta}_{Ai})\\
   &+2S(J_3+\delta J)\sum_i\cos(\bm{k}\bm{\Delta}_{Bi}),\\
   g_{2\bm{k}}=&SJ_2\sum_i\exp(-i\bm{k}\bm{\delta}_i),
\end{aligned}
\end{equation}
and for the V$_{1/3}$NbS$_2$ structure,
\begin{equation}
\begin{aligned}
   g_{a\bm{k}}=&-6SJ_1+6SJ_2-12SJ_3+2SJ_1\sum_i\cos(\bm{k}\bm{d}_i)\\
   &+2S(J_3+\delta J)\sum_i\cos(\bm{k}\bm{\Delta}_{Ai})\\
   &+2S(J_3-\delta J)\sum_i\cos(\bm{k}\bm{\Delta}_{Bi}),\\
   g_{b\bm{k}}=&-6SJ_1+6SJ_2-12SJ_3+2SJ_1\sum_i\cos(\bm{k}\bm{d}_i)\\
   &+2S(J_3-\delta J)\sum_i\cos(\bm{k}\bm{\Delta}_{Ai})\\
   &+2S(J_3+\delta J)\sum_i\cos(\bm{k}\bm{\Delta}_{Bi}),\\
   g_{2\bm{k}}=&SJ_2\sum_i\exp(-i\bm{k}\bm{\delta}_i).
\end{aligned}
\end{equation}

The quadratic Hamiltonian can then be diagonalized by a Bogoliubov transformation, yielding
\begin{equation}
    H_1=E_{0}^{(1)}
    +\sum_{\bm{k}}\Big[E_{+}^{(0)}(\bm{k})\,\alpha_{\bm{k}}^{\dagger}\alpha_{\bm{k}}
    +E_{-}^{(0)}(\bm{k})\,\beta_{\bm{k}}^{\dagger}\beta_{\bm{k}}
    \Big],
\end{equation}
where $E_{0}^{(1)}$ is the order-$S$ quantum correction to the classical ground-state energy, and $E_{\pm}^{(0)}(\bm{k})$ denote the magnon dispersions obtained within LSWT,
\begin{equation}
    E_{\pm}^{\text{op}(0)}(\bm{k})=\sqrt{(g_{\bm{k}}+2SD_z)^2-|g_{2\bm{k}}|^2}\pm \delta g_{\bm{k}},
\end{equation}
\begin{equation}
\begin{aligned}
    E_{\pm}^{\text{ip}(0)}(\bm{k})=&(g_{\bm{k}}^2+\delta g_{\bm{k}}^2-|g_{2\bm{k}}|^2-2SD_zg_{\bm{k}}\\
    &\pm2(\delta g_{\bm{k}}^2((g_{\bm{k}}-SD_z)^2-|g_{2\bm{k}}|^2)\\
    &+S^2D_z^2|g_{2\bm{k}}|^2\big)^{1/2})^{1/2},
\end{aligned}
\end{equation}
where $g_{\bm{k}}=(g_{a\bm{k}}+g_{b\bm{k}})/2$ and $\delta g_{\bm{k}}=(g_{a\bm{k}}-g_{b\bm{k}})/2$.

\bibliographystyle{unsrt}
\bibliography{refs}

\end{document}